\documentclass[twocolumn,tighten,times]{aastex62}

\usepackage{amssymb,amsmath}
\usepackage{graphicx}
\usepackage{color}
\usepackage{subfigure}
\usepackage{lineno}

\newcommand{\oi}{\ion{O}{1}~1355~\AA}

\newcommand{\mgiikk}{\ion{Mg}{2}~k~2796\AA}
\newcommand{\mgiihh}{\ion{Mg}{2}~h~2803\AA}
\newcommand{\mgiik}{\ion{Mg}{2}~k}

\newcommand{\mgiikt}{\ion{Mg}{2}~k3}

\newcommand{\cii}{\ion{C}{2}~1335\AA}

\newcommand{\iris}{{\em IRIS}}

\newcommand{\hmi}{{\em HMI}}

\newcommand{\sst}{{\em SST}}

\submitjournal{ApJ}

\shorttitle{O I 1355\AA\ observations}
\shortauthors{Carlsson \& De Pontieu}

\newcommand{\longacknowledgment}{The authors are grateful to the observers at the \sst\ who obtained the \sst\ data and to Luc Rouppe van der Voort who calibrated and processed the \sst\ timeseries. B.D.P. gratefully acknowledge support by NASA contract NNG09FA40C (\iris). 
This research is supported by the Research Council of Norway through its Centres of Excellence scheme, project number 262622.
 To analyze the data we have used IDL. Data are courtesy of \iris. \iris\ is a NASA small explorer mission developed and operated by LMSAL with mission operations executed at NASA Ames Research Center and major contributions to downlink communications funded by ESA and the Norwegian Space Agency.}

\begin{document}

\title{An optically thin view of the solar chromosphere from observations of the O I 1355\AA\ spectral line}

\correspondingauthor{Mats Carlsson}
\email{mats.carlsson@astro.uio.}

\author[0000-0001-9218-3139]{Mats Carlsson}
\affil{Rosseland Centre for Solar Physics, University of Oslo, P.O. Box 1029 Blindern, NO-0315 Oslo, Norway}
\affil{Institute of Theoretical Astrophysics, University of Oslo, P.O. Box 1029 Blindern, NO-0315 Oslo, Norway}

\author[0000-0002-8370-952X]{Bart De Pontieu}
\affil{Lockheed Martin Solar \& Astrophysics Laboratory, 3251 Hanover Street, Palo Alto, CA 94304, USA}
\affil{Rosseland Centre for Solar Physics, University of Oslo, P.O. Box 1029 Blindern, NO-0315 Oslo, Norway}
\affil{Institute of Theoretical Astrophysics, University of Oslo, P.O. Box 1029 Blindern, NO-0315 Oslo, Norway}

\begin{abstract}
The \ion{O}{1} 1355\AA\ spectral line is one of the only optically thin lines that are both routinely observed and thought to be formed in the chromosphere. We present analysis of a variety of observations of this line with the Interface Region Imaging Spectrograph (\iris), and compare it with other \iris\ diagnostics as well as diagnostics of the photospheric magnetic field. We utilize special deep exposure modes on \iris\ and provide an overview of the statistical properties of this spectral line for several different regions on the Sun. We analyze the spatio-temporal variations of the line intensity, and find that it is often significantly enhanced when and where magnetic flux of opposite polarities cancel. Significant emission occurs in association with chromospheric spicules. Because of the optically thin nature of the \ion{O}{1} line, the non-thermal broadening can provide insight into unresolved small-scale motions.  We find that the non-thermal broadening is modest, with typical values of 5-10 km/s, and shows some center-to-limb variation, with a modest increase towards the limb. The dependence with height of the intensity and line broadening off-limb is compatible with the line broadening being dominated by the superposition of Alfv\'en waves on different structures. The non-thermal broadening shows a modest but significant enhancement above locations that are in between photospheric magnetic flux concentrations in plage, i.e., where the magnetic field is likely to be more inclined with respect to the line-of-sight.  
Our measurements provide strict constraints on future theoretical models of the chromosphere.
\end{abstract}

\keywords{Sun: chromosphere -- Sun: transition region -- Sun: magnetic fields -- magnetohydrodynamics (MHD) }

\section{Introduction}
The solar chromosphere is a highly dynamic and finely structured region of the solar atmosphere that is sandwiched between the visible surface or photosphere and the million-degree outer atmosphere or corona. All non-thermal energy that drives the solar wind and the heating of the corona traverses this critical region. Moreover, despite the only modest enhancement of the chromospheric temperature compared to that of the photosphere, it requires several orders of magnitude more non-thermal energy to drive the dynamics and energetics of the chromosphere than the rest of the solar atmosphere combined. This is because of the high chromospheric densities: the chromosphere contains more mass than the region stretching from the transition region to the edges of the heliosphere. Despite its obvious importance, it is relatively poorly understood, with many open questions remaining regarding the physical processes that drive the dynamics and energetics in the chromosphere \citep{Carlsson2019}. 

\begin{table*}[tph]
\begin{center}
\begin{tabular}{|l|l|c|c|r|r|r|r|c|}
\hline 
Start Date & End Date & Target & OBS-ID & Exp &Solar x & Solar y & $\theta$ & $\mu$\\
\hline
2015-09-09T07:59:58 & 2015-09-09T10:56:08 & AR 12412 & 3610091469 & 15 & -449 & -213 & 32 & 0.82-0.88\\
2015-09-23T00:09:43 & 2015-09-23T02:58:30 & AR 12920 & 3690092077 & 30 & -754 &  104 & 59 & 0.49-0.68\\
2015-09-23T20:22:31 & 2015-09-23T23:11:18 & AR 12920 & 3690092077 & 30 & -565 &   81 & 41 & 0.74-0.85\\
2015-09-24T17:22:13 & 2015-09-24T20:11:00 & AR 12920 & 3690092077 & 30 & -408 &   77 & 29 & 0.86-0.93\\
2015-09-25T21:00:43 & 2015-09-25T23:49:30 & AR 12920 & 3690092077 & 30 & -182 &   63 & 15 & 0.96-0.99\\
2015-09-26T21:23:41 & 2015-09-27T00:12:28 & AR 12920 & 3690092077 & 30 &   27 &   58 &  4 & 0.98-1.00\\
2015-09-28T17:08:41 & 2015-09-28T19:57:28 & AR 12920 & 3690092077 & 30 &  425 &   83 & 23 & 0.84-0.92\\
2015-10-02T00:54:43 & 2015-10-02T03:43:30 & AR 12920 & 3690092077 & 30 &  892 &  140 & 62 & 0.00-0.49\\
2015-10-02T20:47:20 & 2015-10-02T23:36:08 & AR 12920 & 3690092077 & 30 &  917 &  165 & 68 & 0.00-0.50\\
2016-03-04T10:34:35 & 2016-03-04T17:26:20 &    QS    & 3690094078 & 60 &   -3 &  900 & 69 & 0.00-0.55\\
2016-03-05T10:14:03 & 2016-03-05T17:05:48 &    QS    & 3690094078 & 60 &    5 & -890 & 67 & 0.00-0.56\\
2016-03-05T17:17:07 & 2016-03-06T00:08:52 &    QS    & 3690094078 & 60 &    5 & -714 & 48 & 0.55-0.76\\
2016-03-06T00:20:11 & 2016-03-06T07:11:56 &    QS    & 3690094078 & 60 &    5 & -538 & 34 & 0.76-0.89\\
2016-03-06T10:41:36 & 2016-03-06T17:33:21 &    QS    & 3690094078 & 60 &    5 & -890 & 67 & 0.00-0.56\\
2016-03-06T23:01:56 & 2016-03-07T05:53:41 &    QS    & 3690094078 & 60 &   -2 &    7 &  4 & 0.99-1.00\\
2016-03-07T06:05:00 & 2016-03-07T12:56:45 &    QS    & 3690094078 & 60 &   -4 &  728 & 49 & 0.52-0.75\\
\hline
\end{tabular}
\caption{For all IRIS observations used, we provide the start and end date, target, IRIS OBS-ID, exposure time (s), solar x/y coordinates (\arcsec), $\theta$ (the viewing angle between the local vertical and line-of-sight vector, in degrees), and $\mu = \cos \theta$.}
\label{table1}
\end{center}
\end{table*}

One of the main reasons for our limited knowledge is the lack of unambiguous diagnostics. Most spectral lines that emanate from the chromosphere are optically thick and are subject to non-LTE radiative transfer effects such as scattering, partial frequency redistribution, etc \citep[e.g.,][]{Leenaarts2013a,Leenaarts2013b}. In addition, non-equilibrium ionization plays a key role in the chromosphere and for some chromospheric diagnostics \citep[e.g.,][]{2002ApJ...572..626C,Golding:2014fk,Golding2016, Leenaarts2016, Leenaarts2017}. This renders their interpretation difficult and dependent on inversion techniques that often suffer from limiting assumptions and/or non-uniqueness, despite major advances in techniques \citep{de-la-Cruz-Rodriguez2016,Sainz-Dalda2019}.
One of the few spectral lines that is optically thin and formed in the chromosphere is the \ion{O}{1} 1355.598~\AA\ intersystem line 
($2s^2\,2p^3\,3s\,^5\!S_2 - 2s^2\,2p^4\,^3\!P_2 $), hereafter called the \ion{O}{1} 1355\AA\ line. It is routinely observed with the Interface Region Imaging Spectrograph \citep[\iris,][]{De-Pontieu2014a}. \citet{Lin2015} performed an analysis of the formation of this line in an advanced numerical simulation calculated with the Bifrost code \citep{Gudiksen:2011qy} and demonstrated that the line formation is optically thin. 

In this paper we provide an overview of observational findings with \iris\ related to the \ion{O}{1} 1355\AA\ line. We describe the observations and analysis techniques of the spectral line profiles in Section~\ref{obs}. We then describe the statistical properties and center-to-limb variation of the intensity and line broadening for various solar targets in Section~\ref{stats}. We also describe how the properties of \ion{O}{1} 1355\AA\ are impacted by flux cancellation and the presence of flux concentrations in, respectively, Sections~\ref{cancel} and \ref{tube}. We finish with a  discussion and conclusions in Section~\ref{dis}.

\section{Observations and analysis techniques}\label{obs}

For the observational analysis we used several different datasets from \iris. For all of these observations we chose to use OBS-ID 3610091469 (for NOAA AR 12412), 
OBS-ID 3690092077 (for AR 12920) or 3690084078 (for quiet Sun), in order to maximize the signal-to-noise (S/N) ratio while maintaining high spectral resolution (to resolve the narrow \ion{O}{1} 1355\AA\ line) by introducing long exposures (15s for
AR 12412, 30s for AR 12920, 60s for QS), lossless compression, and asymmetric summing (x2 summing spatially, no summing spectrally). AR 12920 was followed over multiple days in September 2015 (see Table~\ref{table1}), as the AR crossed the disk. The field-of-view of the observations was 
45\arcsec x 120\arcsec (AR 12412), 112\arcsec x 175\arcsec (AR 12920) and 140\arcsec x 175\arcsec (QS) with a spatial sampling of 0.35\arcsec\ by 0.33\arcsec. The full detector was read out, i.e., three different wavelength ranges covering wavelengths from, respectively, 1331.56–1358.40\AA\ (FUV1), 1390.00–1406.79\AA\ (FUV2), and 2782.56–2833.89\AA\ (NUV), with a spectral sampling of 12.98 m\AA\ in FUV1, 12.72 m\AA\ in FUV2, and 25.46 m\AA\ in NUV. The nominal spatial resolution of \iris\ is 0.33\arcsec\ in the FUV and 0.4\arcsec\ in the NUV. Note that since we used spatially summed data, our spatial resolution along the slit is Nyquist limited to twice the plate scale (0.66\arcsec, i.e., 2 times 0.33\arcsec). The spectral resolution (FWHM) of \iris\ is dominated by Nyquist sampling, i.e., 5.7 km/s in both FUV and NUV.

To determine the first moments of the \ion{O}{1} spectral line, we used the following steps: 1. remove the impact of cosmic rays on the detector (using the {\tt clean\_exposure.pro} code in IDL SolarSoft), 
2. remove the impact of fixed pattern noise in an extra dark current subtraction,
3. take the mean of 3x3 pixels to increase the S/N.
4. apply a single Gaussian fit to the spectral line profile in each location. 

The second step of this procedure is non-standard and is therefore described in some detail here. The \ion{O}{1} 1355\AA\ line is rather weak and especially the width determination is influenced by errors in the dark-level. We often see horizontal stripes of increased width in the width-maps. These stripes are in the same location over some time and are caused by fixed patterns in the dark-level that are not removed in the standard reduction pipeline. When a "hot" pixel is present at a location in the wing of the \ion{O}{1} line, a larger width results from the Gaussian fit. We used the QS dataset from the center of the disk (2016-03-06T23:01:56) and formed a mean intensity, $I(y,\lambda)$ by taking the mean over scan position x of $I(x,y,\lambda)$ for all positions
where the line-core intensity was less than 15 DN. This mean will still have an imprint of the spectral line. We removed that by subtracting at each wavelength and pixel y the mean over $\pm$ 10 pixels. The fixed pattern noise is not completely fixed in time so this procedure makes the most significant improvement to the width determinations 
for the QS datasets, and not so much for the AR 12920 datasets. The time-scales on which the fixed pattern changes significantly appear to be days to months. This issue and potential fixes in the pipeline are currently being studied in more detail by the IRIS calibration team.

The single Gaussian fit is made with three free parameters: the maximum intensity, the Doppler shift of the profile and the width. A constant background is first determined from a mean of the spectrum at blue-shifts of between 200 and 40 km/s relative to the rest-wavelength of the \ion{O}{1} 1355\AA\ line (a line-free region of the spectrum). This constant background is subtracted from the spectrum before the single Gaussian fit. The fit is only including the spectrum between -40 and 40 km/s relative to the rest-wavelength of the \oi\ line, in order to avoid influence from the \ion{C}{1} line at 1355.843 \AA\ (corresponding to 54 km/s from the rest-wavelength of the \oi\ line). A fit is attempted for all pixels but the fit is judged unsuccessful if the fitting algorithm ({\tt mpfitpeak.pro} in IDL SolarSoft) does not converge or the determined values are outside reasonable values ([0,1000] DN in maximum intensity, [-30,30] km/s in Doppler shift, [2,30] km/s in width).

The reported line broadening is the 1/e half-width (henceforth referred to as the 1/e width) as this provides the most probable velocity, which is useful for interpretation in terms of physical mass motions in the solar atmosphere. The line width has several contributions: $w= \sqrt{w^2_{inst}+w^2_{th}+w^2_{nth}}$. We note that the instrumental broadening $w_{inst}$ in the FUV channel of \iris\ has been reported \citep{De-Pontieu2014a} as 25.85 m\AA\ (i.e., 5.7 km/s) full-width-half-maximum (FWHM), which translates to 15.52 m\AA\ or 3.4 km/s 1/e width. The thermal broadening $w_{th}$ of the \oi\ line is somewhat uncertain given the large range of chromospheric heights this line is formed over. If we assume that the chromospheric temperature is within a range of 5,000 to 15,000~K, the thermal broadening would range between 2.3 and 3.9 km/s (1/e width). Both the instrumental and thermal broadening are smaller than most widths we report on here and play only a minor role. For the \ion{Mg}{2} k profiles we used a double Gaussian fit similar to that used by \citet{Carlsson2015} and \citet{Bryans2016}. The reported width is the 1/e width of the Gaussian fit to the profile outside the central reversal. For the \ion{C}{2} profiles we used a double Gaussian fit similar to that used by \citet{Rathore2015b}.

For some \iris\ observations we also provide magnetogram data from \hmi\ for context. The magnetogram data is based on the circular polarization Stokes signal in the \ion{Fe}{1} 6173\AA\ line and provides information on the line-of-sight (LOS) magnetic field at photospheric levels. The \hmi\ data has a pixel size of 0.5\arcsec. We use the full-disk data to obtain a field-of-view that is matched to that of \iris. The co-alignment of the \iris\ and \hmi\ magnetogram data is accomplished through comparison of an \iris\ spectroheliogram at 2800\AA\ in the photospheric wings of the \ion{Mg}{2} h \& k lines and the \hmi\ magnetogram. This alignment is greatly facilitated by the similarities between the bright points in the 2800\AA\ images and the flux concentrations that are detected with \hmi.

For one of the \iris\ observations, we also provide high-resolution context magnetograms of the circular polarization in the \ion{Fe}{1} 6173\AA\ line using the CRISP instrument \citep{Scharmer:2006gf} at the Swedish 1-m Solar Telescope (\sst). This data was obtained on 2015-09-09 from 09:04:12 UTC to 10:29:04 UTC with a cadence of 67 seconds. The pixel size is 0.06\arcsec. The data was calibrated and corrected for deformations because of seeing conditions introduced by the Earth's atmosphere in an identical manner as described in \S~2.2 in \citet{Rouppe-van-der-Voort2020}.

\begin{figure*}
    \centering
    \includegraphics[width=0.99\textwidth]{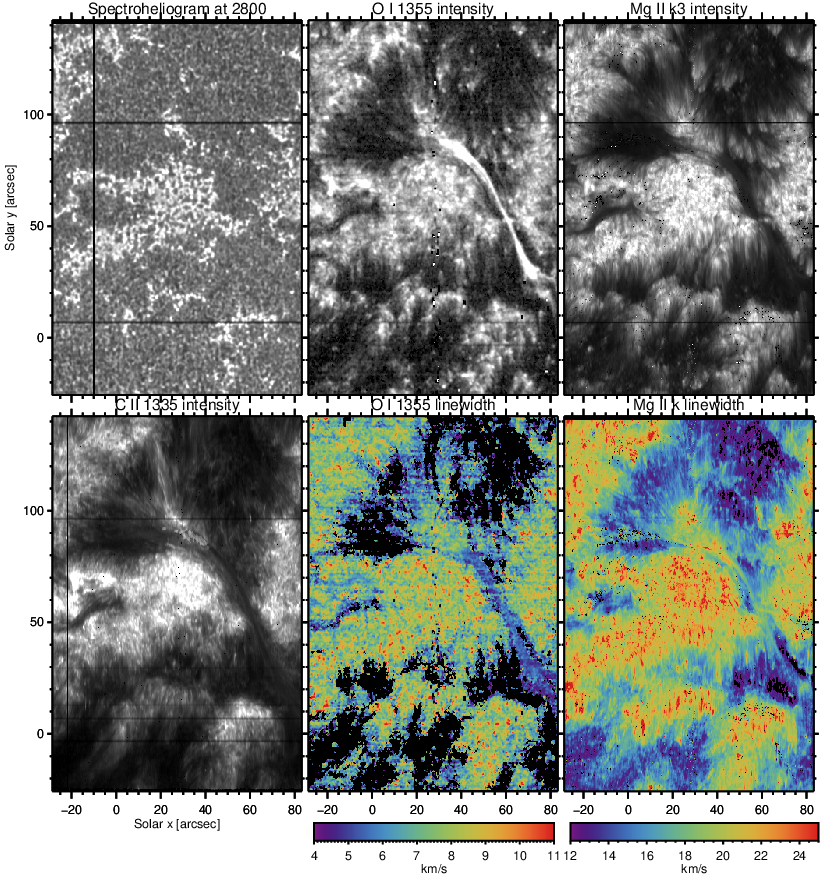}
    \caption{\iris\ spectroheliograms of NOAA AR 12920 at 2015-09-26T21:23 UTC showing in the top row 2800\AA, \ion{O}{1} 1355\AA\ intensity, and \ion{Mg}{2} k3 intensity, and in the bottom row \ion{C}{2} 1335\AA\ intensity, \ion{O}{1} 1355\AA\ line broadening, and the \ion{Mg}{2} k line broadening. Black horizontal lines are fiducial marks. Black vertical lines are data drop-outs. The \oi\ intensity, \mgiikt\ intensity, and \ion{C}{2} 1335\AA\ intensity are scaled, respectively, between 1 and 35 DN, 0 and 10586 DN, and 3 and 311 DN. Regions with bad fits and/or low peak counts (below 6 DN) are masked out in the \oi\ line width map. This figure is accompanied by an animation that allows the reader to blink between the different panels of the figure to see the various similarities and offsets described in the text.}
    \label{fig:overview}
\end{figure*}

\section{Properties of \oi} \label{stats}

\subsection{Morphology and relationship to other spectral lines}

\citet{Lin2015} used numerical simulations of a region that aims to mimic an enhanced network region in quiet Sun and found that the line forms through a recombination cascade to the upper level of the \ion{O}{1} 1355\AA\ line, followed by emission in the line under optically thin conditions. The ionization balance of \ion{O}{1} and \ion{O}{2} is coupled to the ionization of hydrogen through charge exchange. The \oi\ line is thus formed in the chromosphere, as \ion{O}{2} ionization occurs towards the top of the chromosphere. Since the line is optically thin, the line broadening is of particular interest as it provides constraints on turbulent motions in the chromosphere, a property otherwise only accessible through spectral line inversions of optically thick chromospheric lines such as \mgiihh\ and \mgiikk\ \citep[e.g.,][]{de-la-Cruz-Rodriguez2016}. Since the line is chromospheric, it is of interest to study the morphological properties of spectroheliograms or maps of the \oi\ intensity and width.

Figure~\ref{fig:overview} shows that in an active region at disk center, the \oi\ line is bright in plage regions and some parts of filaments. The morphology appears to be a combination of dot-like and more extended features. The regions where \oi\ is bright (top row, middle panel) slightly extend beyond the plage perimeters that are seen in a photospheric 2800\AA\ spectroheliogram (top row, left panel). The extensions seem wispy and resemble the lower parts of loop-like or fibril-like features. 

There is overall a strong resemblance with the \mgiikt\ intensity maps\footnote{The \mgiikt\ spectral feature occurs at the wavelength of the central reversal for a double-peaked \mgiik\ spectral profile, or the wavelength of the peak for a single-peaked \mgiik\ spectral profile \citep{Leenaarts2013a}}. The main difference between both maps is that the regions where \mgiikt\ is bright extend even further beyond the photospheric boundaries of the plage, possibly because the signal-to-noise (S/N) of the \oi\ line is significantly lower. The other difference is that filaments are typically dark in \mgiikt\ and most often bright in \oi\ (but see the filament around $x=-10$\arcsec, $y=60$\arcsec\ which is dark in both). There are also significant morphological similarities with the \cii\ intensity (bottom row, left panel). The main difference here is that the latter shows more differences in brightness between various plage regions, because it is more sensitive to transition region conditions \citep{Rathore2015a}, which are, in turn, dominated by the overlying coronal conditions.  In summary, our comparison shows that in plage regions the \oi\ line does indeed look very chromospheric in nature with significant similarity between the intensity patterns in the \oi\ and \mgiik\ line. 

\begin{figure}
    \centering
    \includegraphics[width=0.45\textwidth]{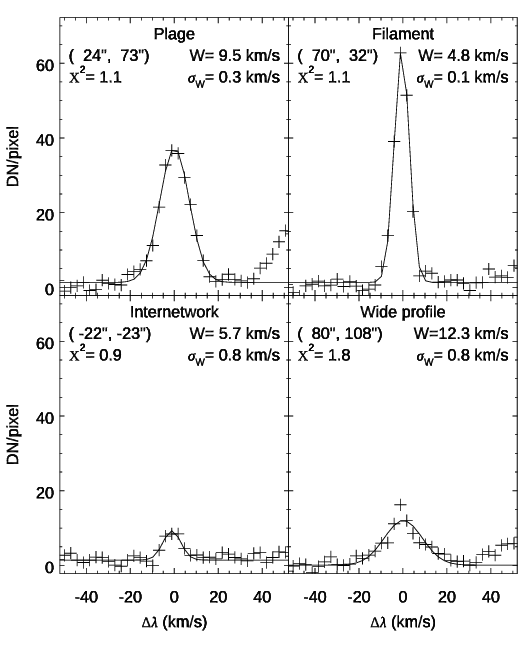}
    \caption{Examples of \oi\ profiles from four locations of Fig.~\ref{fig:overview}. The observed profile is given with symbols and the single Gaussian fit with a solid line. The positions and reduced $\chi^2$ of the fit are given in the upper left with the 1/e width and its 1$\sigma$ error in the upper right. The increased intensity in the red part of the window is due to an emission line from \ion{C}{1} at 1355.8 \AA.}
    \label{fig:ifit}
\end{figure}

\begin{figure*}[tph]
    \centering
    \includegraphics[width=0.99\textwidth]{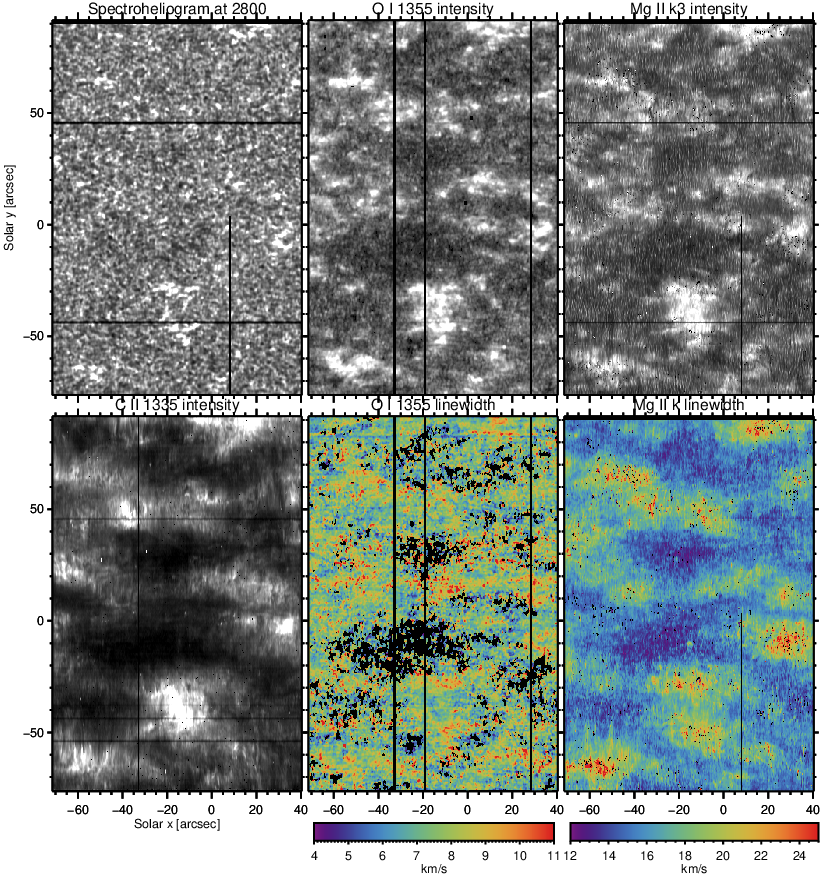}
    \caption{\iris\ spectroheliograms of a quiet Sun region at 2016-03-06T23:01 UTC showing in the top row 2800\AA, \ion{O}{1} 1355\AA\ intensity, and \ion{Mg}{2} k3 intensity, and in the bottom row \ion{C}{2} 1335\AA\ intensity, \ion{O}{1} 1355\AA\ line broadening, and the \ion{Mg}{2} k line broadening. Black horizontal lines are fiducial marks. Black vertical lines are data drop-outs. The \oi\ intensity, \mgiikt\ intensity, and \ion{C}{2} 1335\AA\ intensity are scaled, respectively, between 2 and 141 DN, 0 and 9563 DN, and 19 and 125 DN. Regions with bad fits and/or low peak counts (below 8 DN) are masked out in the \oi\ line width map. This figure is accompanied by an animation that allows the reader to blink between the different panels of the figure to see the various similarities and offsets described in the text.}
    \label{fig:overview_qs}
\end{figure*}

The spatial patterns in the map of \oi\ linewidths (bottom row, middle panel) generally map, on large scales, those of the \oi\ intensity (top row, middle panel). The \oi\ linewidth is enhanced in and around plage regions, and very low in surrounding quiet Sun regions. The map of \oi\ line widths is remarkably homogeneous within each plage region, with only relatively small deviations around the average value of about 7.5 km/s, as previously remarked by \citet{Carlsson2015}. For each plage region, the high values of \oi\ widths spatially extend even further beyond the photospheric plage boundaries than the \oi\ intensity. The values of \oi\ broadening in quiet Sun are very small, often insignificant, suggesting that, in many quiet Sun locations, the line is not broadened beyond the combination of instrumental and thermal broadening.  

The spatial patterns of the line broadening in \oi\ and \mgiik\ (bottom row, right panel) show significant correspondence on very large scales ($\sim$10\arcsec): both lines are broader in plage regions and the immediate vicinity, and narrower in the quiet Sun regions surrounding the plage. However, the broadening values themselves are very different between these two lines. Typical values for \oi\ are of order 10 km/s or less, while the \mgiik\ values are larger by a factor of $\sim$3. This is in agreement with the results from \citet{Carlsson2015}. It is not surprising given that \oi\ is an optically thin line and its width is sensitive to velocity variations along the LOS and turbulent motions, while \mgiik\ is an optically thick line with velocity gradients, turbulent motions, and broadening as a result of opacity all playing a role in the line width. On very small ($< 1$\arcsec) scales there is most often not a good match between \oi\ broadening and \mgiik\ broadening. 

Figure~\ref{fig:ifit} gives four examples of observed profiles together with the single Gaussian fit of the \oi\ line from four different regions seen in Fig.~\ref{fig:overview}: Plage, a filament, internetwork and an unusually wide profile. Note that the spectra have been averaged over 3x3 spatial pixels -- the actual number of photons per wavelength bin is thus nine times the given DN/pixel times four photons per DN \citep[the gain in the FUV passband\footnote{The gain in the NUV passband is 18 photons per DN.},][]{De-Pontieu2014a}. The dominant error source is photon-noise, proportional to the square-root of the number of photons for Poisson statistics. For very low count-rates, a number of other error sources come into play \citep[readout-, digitization-, flat-fielding, fixed-pattern-noise, straylight subtraction, etc, ][]{Wulser2018}. We estimate the total error as the square-root of the sum of the squares of a term not dependent on the count-rate (estimated from
the standard deviation of the count-rate in a line-free region between -200 and -40 km/s) and the Poisson noise term. We furthermore restrict our analysis to profiles with a line-centre count-rate above a threshold (given in the respective figures). For the three first profiles, the fit with a single Gaussian is very good. The unusually wide profile has a markedly non-Gaussian shape. This is typical for the widest profiles.

In quiet Sun, we find many of the same properties and correspondences, as shown in Fig.~\ref{fig:overview_qs}. The \oi\ is bright in and around quiet Sun network regions and  faint in the internetwork regions. At disk center, we see that the \oi\ line is too weak to detect in some of the internetwork regions, despite the deep exposures and spatial summing in the special observing mode for these observations. However, in much of the FOV there is sufficient signal to determine the intensity and line width of the \oi\ line. The findings for QS are very similar to those described above for active regions.
The strongest signal can be found in and around network regions. A disk center view shows that the regions where \oi\ intensity is high again extend beyond the spatial boundaries of the network, similar to what we see in plage regions. The network-associated \oi\ emission appears to be a combination of dot-like features and more extended wispy structure that are reminiscent of the spicules that are often seen to protrude from network flux concentrations. The internetwork regions are significantly fainter. The \oi\ intensity map shows strong similarity with the \mgiikt\ intensity map. There is, again, similarity with the \cii\ intensity map, although somewhat less than with \mgiikt, with \cii\ showing higher contrast and strongly reduced internetwork intensities.

The \oi\ line width maps reveal extremely low values in the internetwork regions. In many locations these values are so low that they appear compatible with a lack of non-thermal broadening. The network regions show a significant increase in the line width, with the regions of high broadening extending even further from the network regions than the intensity itself. This is similar to what we found for plage, although the line width is somewhat reduced in network compared to plage. The \oi\ broadening is much less than that found in the \mgiik\ line, again similar to what we found for active regions.

\begin{figure*}
    \centering
    \includegraphics[width=0.85\textwidth]{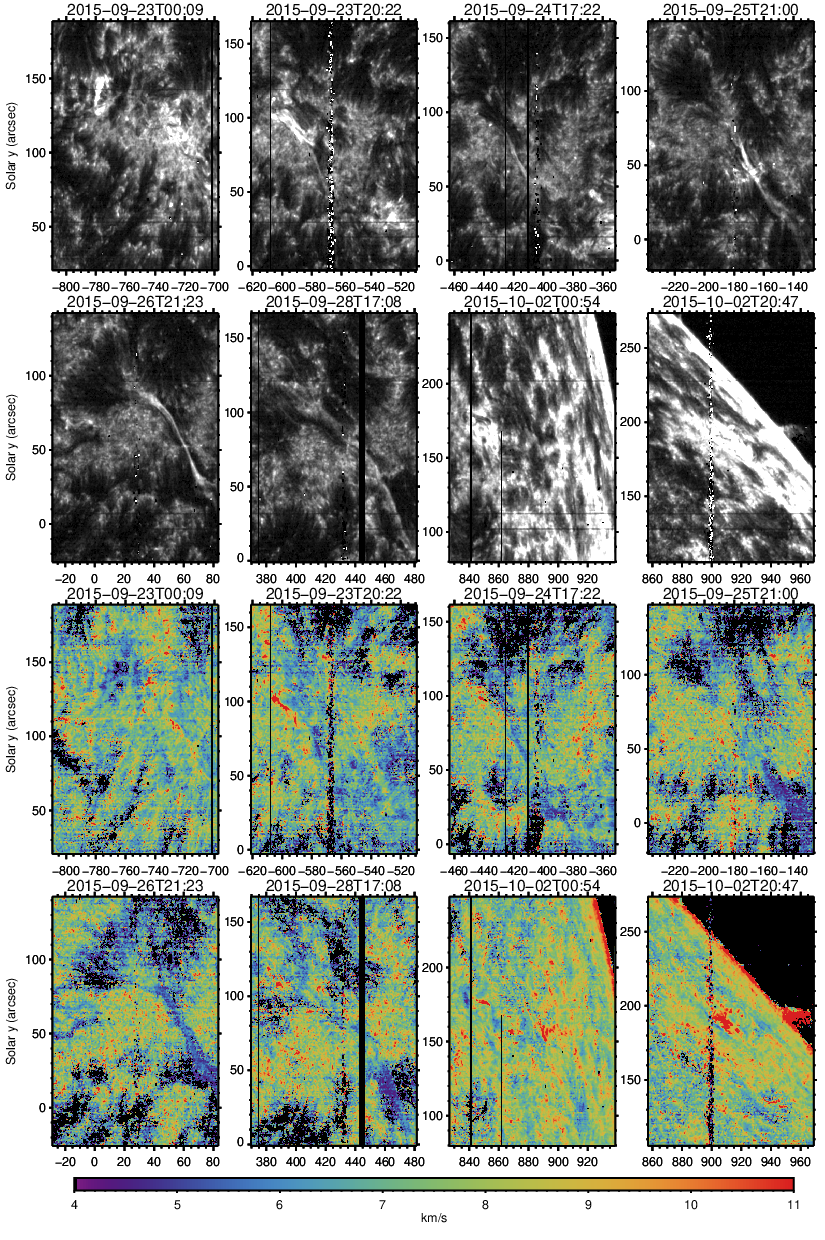}
    \caption{\iris\ spectroheliograms of NOAA AR 12920 as it traverses the disk between 2015-09-22 and 2015-10-02. Top two rows show the  \ion{O}{1} 1355\AA\ intensity, while the bottom two rows show the \ion{O}{1} 1355\AA\ line broadening. The \oi\ intensity is scaled between 0 and 60 DN for all panels. Regions with bad fits and/or low peak counts (below 4 DN) are masked out in the \oi\ line width maps.}
    \label{fig:c2l_ar}
\end{figure*}

To gain further insight into the morphology and formation region of \oi\ it is of interest to study the appearance of the \oi\ intensity and line width in solar targets that are closer to the limb, both for an active region and a quiet Sun region. Fig.~\ref{fig:c2l_ar} shows NOAA AR 12920 as it is tracked over the course of 9 days from close to the east limb to the west limb of the Sun, with both the minimum and maximum of the color scale for both intensity and line width identical for all images. It appears that the intensity is lowest close to disk center, with a steady increase as we approach the limb. At the limb itself the intensity is increased significantly, as is expected from an optically thin line: the line-of-sight increases and captures more structures as the viewing angle is more oblique. We also see localized regions of strong intensity enhancements, in particular on 2015-09-23, 2015-09-24, 2015-09-25, and 2015-10-02. We will describe these very bright regions in more detail in \S~\ref{cancel}.

\begin{figure*}
    \centering
    \includegraphics[width=0.99\textwidth]{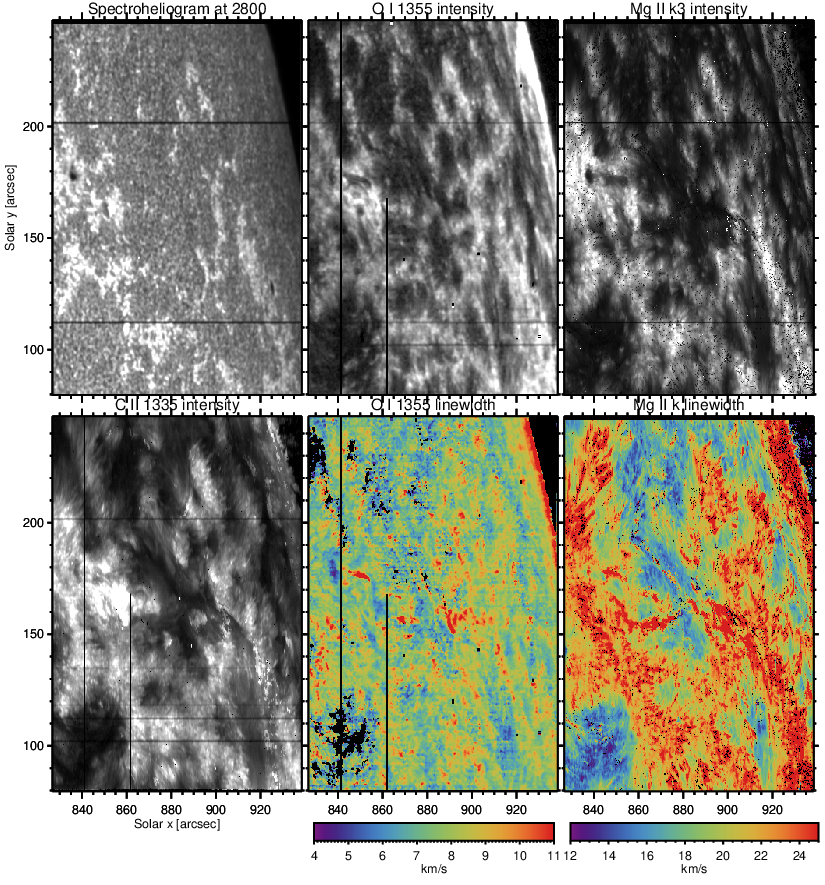}
    \caption{\iris\ spectroheliograms of NOAA AR 12920 at 2015-10-02T00:54 UTC showing in the top row 2800\AA, \ion{O}{1} 1355\AA\ intensity, and \ion{Mg}{2} k3 intensity, and in the bottom row \ion{C}{2} 1335\AA\ intensity, \ion{O}{1} 1355\AA\ line broadening, and the \ion{Mg}{2} k line broadening. Black horizontal lines are fiducial marks. Black vertical lines are data drop-outs. The \oi\ intensity, \mgiikt\ intensity, and \ion{C}{2} 1335\AA\ intensity are scaled, respectively, between 0 and 92 DN, 0 and 13169 DN, and 3 and 224 DN. Regions with bad fits and/or low peak counts (below 6 DN) are masked out in the \oi\ line width map. This figure is accompanied by an animation that allows the reader to blink between the different panels of the figure to see the various similarities and offsets described in the text.}
    \label{fig:ar_limb}
\end{figure*}

The line width remains relatively independent of the viewing angle except there is a clear increase for viewing angles close the limb (e.g., 2015-09-23, 2015-10-02). At the limb itself the line width increases in a step-like fashion, which is to be expected since the integration length doubles at the limb. Nevertheless the line widths remain relatively modest and the overall variation is limited to just a few km/s on average, as described below using histograms.

\begin{figure*}
    \centering
    \includegraphics[width=0.99\textwidth]{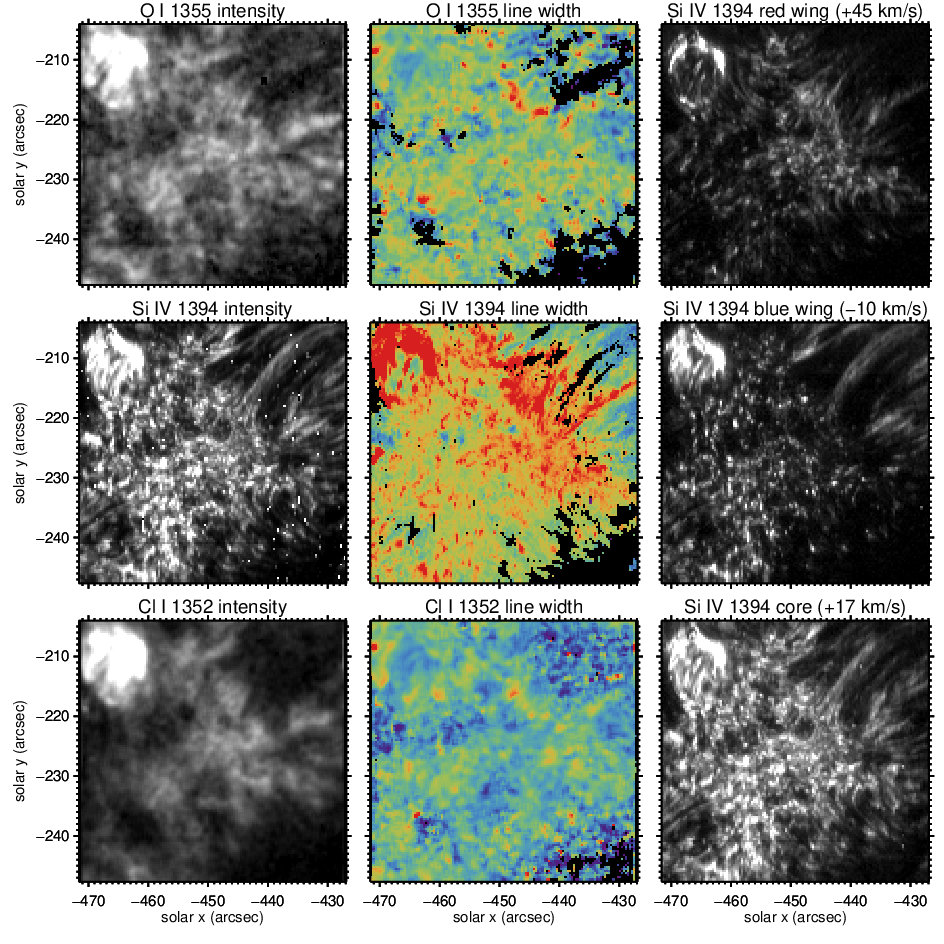}
    \caption{\iris\ views of NOAA AR 12412 at 2015-09-09T07:59 UTC. The right column shows spectroheliograms of \ion{Si}{4} 1394\AA\ in the red wing (top), blue wing (middle), and core (bottom). The left and middle columns show, respectively the intensity and line broadening from a Gaussian fit for the \oi\ line (top), \ion{Si}{4} 1394\AA\ (middle), and \ion{Cl}{1} 1351\AA\ (bottom). The color scale for the line broadening is between 3 and 12 km/s for \ion{O}{1} and \ion{Cl}{1}, and between 5 and 40 km/s for \ion{Si}{4}. The \oi\ intensity, \ion{Si}{4} red wing intensity, \ion{Si}{4} intensity, \ion{Si}{4} blue wing intensity, \ion{Cl}{1} intensity, and \ion{Si}{4} core intensity, are scaled, respectively, between 2 and 30 DN, 5 and 500 DN, 10 and 500 DN, 5 and 500 DN, 10 and 250 DN, and 5 and 500 DN. Regions with bad fits and/or low peak counts (below 5 DN) are masked out in the \oi\ line width map. This figure is accompanied by an animation that allows the reader to blink between the different panels of the figure to see the various similarities and offsets described in the text.} 
    \label{fig:ar_blink}
\end{figure*}

A more detailed view of NOAA AR 12920 on 2015-10-02 is provided in Fig.~\ref{fig:ar_limb}. At this time the AR is close to the limb so that the FOV covers a wide range of values for $\mu = \cos \theta$. At this extreme viewing angle it appears that \oi\ intensity features (top row, middle panel) are slighly offset towards the limb when compared with the photospheric plage footpoints (top row, left panel). In addition, the \oi\ intensity features that protrude towards the limb from the photospheric footpoints protrude less (i.e., are shorter) than the features in the \mgiikt\ intensity maps. The protrusions are even more extended towards the limb in the \cii\ images. They are strongly reminiscent of spicules. This suggests that the \oi\ line is indeed formed in the chromosphere, but perhaps at lower heights than the upper chromospheric and lower transition region features that are visible in the \mgiikt\ and \cii\ maps. This difference in apparent formation height is perhaps even more clear when comparing the line width maps in \oi\ and \mgiik. The regions with enhanced \oi\ line width around the plage regions are relatively narrow, and definitely shorter than the equivalent features in \mgiik. In addition, the \oi\ line width map appears to show significantly less enhancement (compared to the \mgiik\ linewidth map) and the region where the line width is enhanced covers a smaller part of the plage regions.

A comparison between \ion{Si}{4} 1402, \ion{Cl}{1} 1352\AA\ and \oi\ spectroheliograms of NOAA AR 12920 towards the limb is also illustrative (Fig.~\ref{fig:ar_blink}).  The \ion{Cl}{1} line intensity\footnote{The moments of the \ion{Cl}{1} line are calculated in the same way as for \ion{O}{1}.}, thought to be formed at similar heights in the chromosphere \citep{1983ApJ...266..882S}, shows a large degree of similarity to that of \oi. That correlation is also present to some extent for the line width of both lines. 
For the \ion{Si}{4} this is different. Here we see a clear difference between the limbward side and the disk-center side of the plage region. Towards the limb there is a clear offset of order 1-2\arcsec, with the \ion{Si}{4} features offset towards the limb. The \ion{Si}{4} brightenings are quite well understood \citep{Skogsrud2016}, and are the TR counterparts to the magneto-acoustic shocks that dominate the plage chromosphere, as can be determined from $\lambda-t$ plots. This comparison suggests that the \oi\ intensity features occur at lower heights than the \ion{Si}{4} shocks. Such shocks are visible both in the blue and red wing of the \ion{Si}{4} line, with the blue wing showing the upward phase and the red wing the downward phase. These shocks have been shown to drive dynamic fibrils \citep{Skogsrud2016} and type II spicules \citep{Rouppe-van-der-Voort2015}. It seems that many \oi\ intensity features are somehow related to both, with the small round features possibly related to shocks (as seen from above), and the wispy linear features possibly related to type II spicules (when viewed from the side). 

\begin{figure*}
    \centering
    \includegraphics[width=0.85\textwidth]{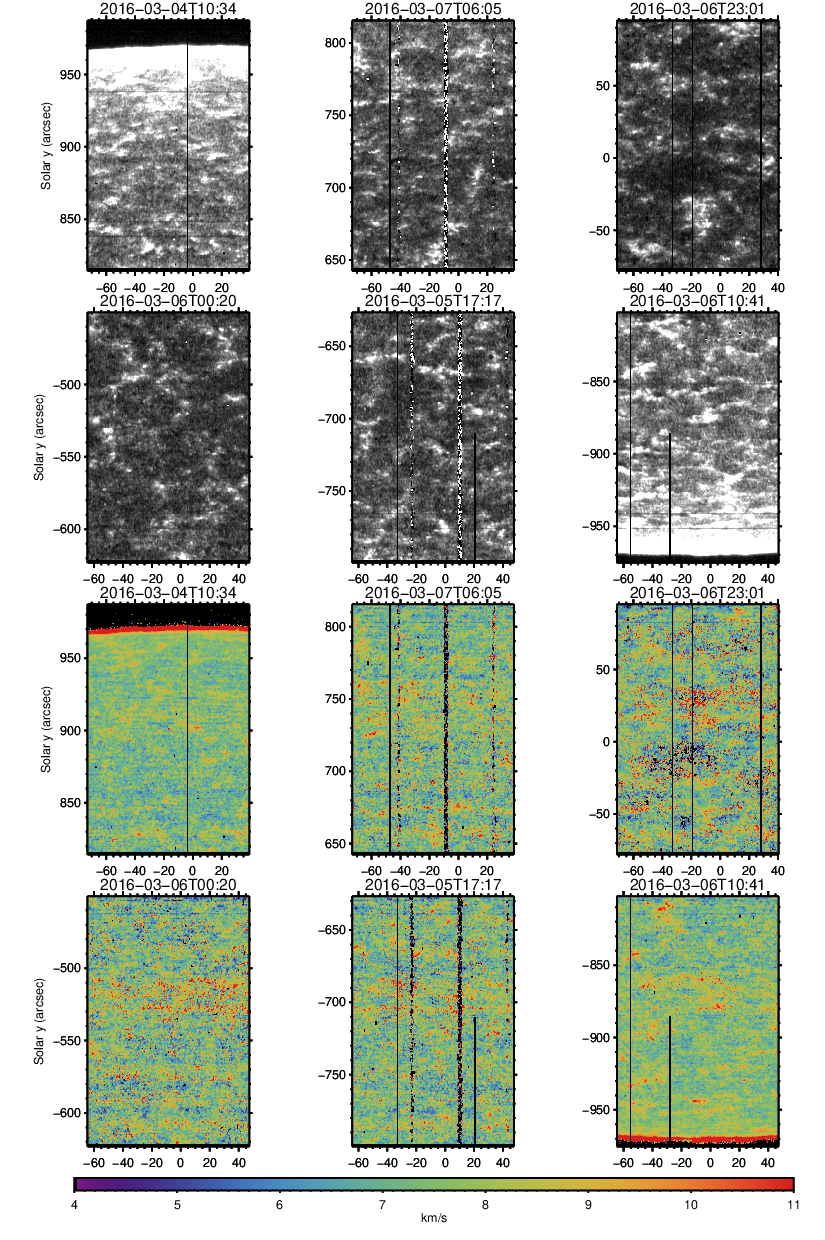}
    \caption{\iris\ spectroheliograms of quiet Sun taken along the meridian from the north pole (top left) to the south pole (bottom right). Top two rows show the  \ion{O}{1} 1355\AA\ intensity, while the bottom two rows show the \ion{O}{1} 1355\AA\ line broadening. The \oi\ intensity is scaled between 0 and 60 DN for all panels. Regions with bad fits and/or low peak counts (below 4 DN) are masked out in the \oi\ line width maps.}
    \label{fig:c2l_qs}
\end{figure*}

The behavior of the \oi\ line in quiet Sun regions appears to be similar. We used observations that were obtained along the central meridian in March 2016 covering several locations between the north and the south pole, as shown in Fig.~\ref{fig:c2l_qs}. We see very similar behavior, with the intensity lowest around disk center, and significantly increased at or close to the limb, again as expected from an optically thin line. The line width is enhanced around the network regions, independent of the viewing angle, and lowest at disk center. The increase of the line width exactly at the limb is significant but only a few km/s. There is also a more gradual increase of the line width from disk center towards the limb, but only by about 1 or 2 km/s.

\begin{figure*}[tph]
    \centering
    \includegraphics[width=0.99\textwidth]{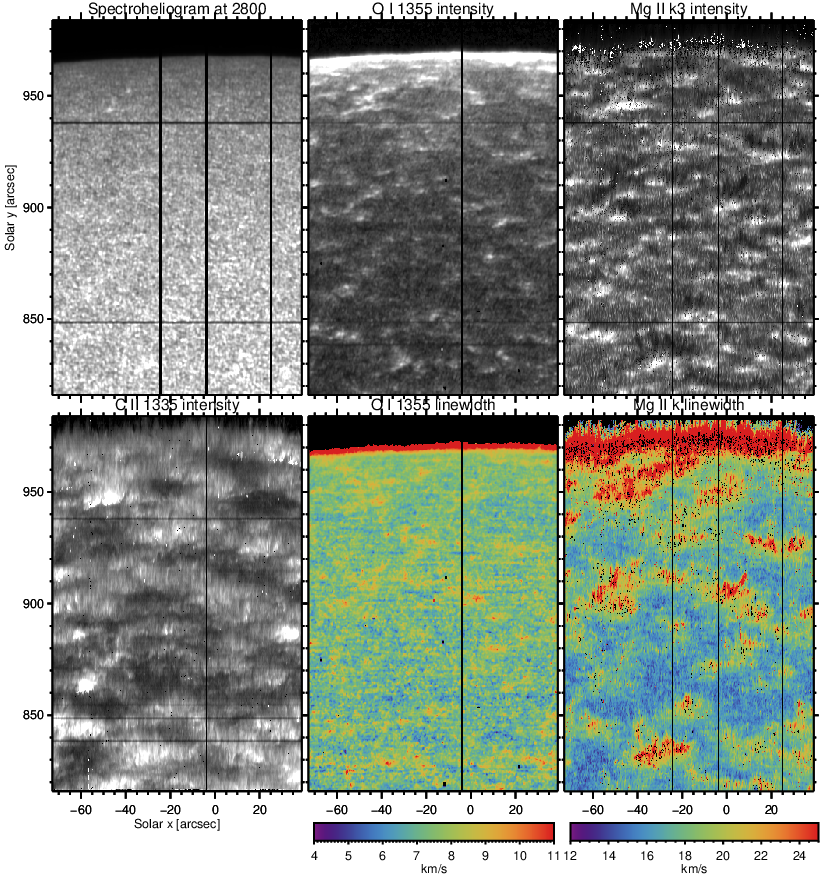}
    \caption{\iris\ spectroheliograms of a quiet Sun at 2016-03-06T23:01 UTC showing in the top row 2800\AA, \ion{O}{1} 1355\AA\ intensity, and \ion{Mg}{2} k3 intensity, and in the bottom row \ion{C}{2} 1335\AA\ intensity, \ion{O}{1} 1355\AA\ line broadening, and the \ion{Mg}{2} k line broadening. Black horizontal lines are fiducial marks. Black vertical lines are data drop-outs. The \oi\ intensity, \mgiikt\ intensity, and \ion{C}{2} 1335\AA\ intensity are scaled, respectively, between 0 and 150 DN, 0 and 7865 DN, and 4 and 108 DN. Regions with bad fits and/or low peak counts (below 8 DN) are masked out in the \oi\ line width map. This figure is accompanied by an animation that allows the reader to blink between the different panels of the figure to see the various similarities and offsets described in the text.}
    \label{fig:overview_qs_limb}
\end{figure*}

A detailed view at the polar limb again shows an offset towards the limb of the bright \oi\ intensity features when comparing to the photospheric network concentrations (visible in 2800\AA). The \oi\ intensity features appear to be shorter (in the direction towards the limb) and offset below the features in \mgiik3\ (upper chromosphere) and \cii\ (TR).

A coherent picture then appears in which, for both active regions and quiet Sun, the \oi\ formation height appears to be chromospheric in nature, but somewhat lower and thus not quite identical to that of \mgiik.

\begin{figure}[tph]
    \centering
    \includegraphics[width=0.45\textwidth]{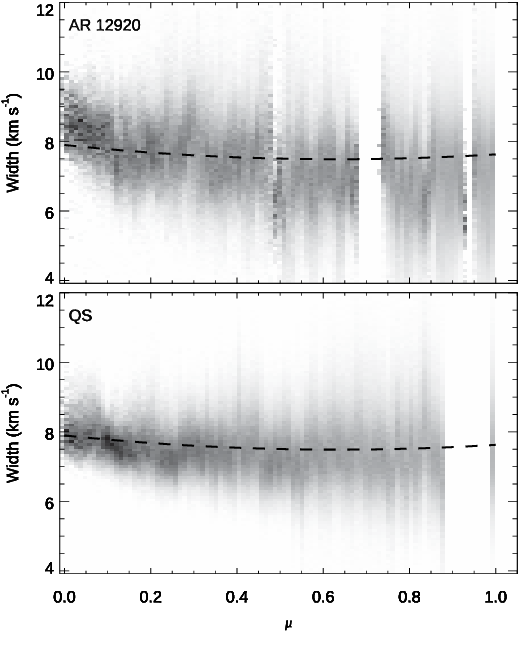}
    \caption{Probability density function (PDF) of the \oi\ line width as function of the cosine of the viewing angle ($\mu$) for the AR 12290 datasets (top panel) and the QS datasets (bottom panel). To facilitate comparison, a quadratic least-squares fit to the QS width as function of $\mu$ is shown in both panels as a dashed line.
    The color-scale is linear and profiles with a central intensity below 10 DN/pixel are not included.}
    \label{fig:wmu_ar_qs}
\end{figure}

\begin{figure}[tph]
    \centering
    \includegraphics[width=0.45\textwidth]{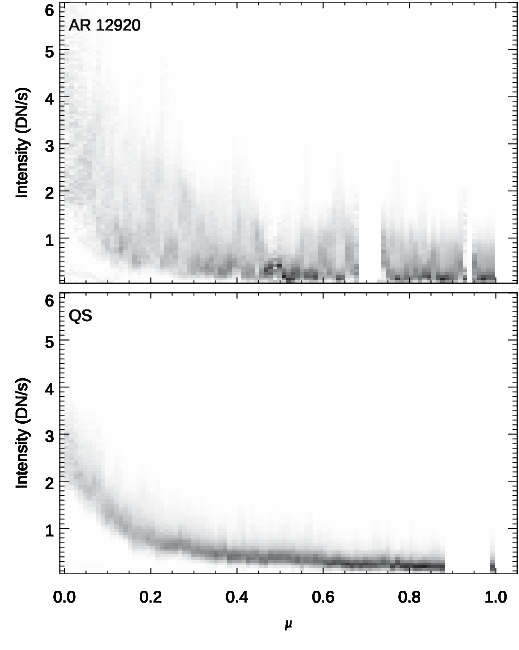}
    \caption{Probability density function (PDF) of the \oi\ line core intensity (from the fit) as function of the cosine of the viewing angle ($\mu$) for the AR 12290 datasets (top panel) and the QS datasets (bottom panel).
    The color-scale is linear}
    \label{fig:imu_ar_qs}
\end{figure}

\subsection{Statistics and center to limb variation} \label{sub:stats}

 Since the \oi\ line is optically thin in the chromosphere, it is of interest to study how the line broadening depends on the viewing angle. The probability density function (PDF) of the line width as a function of the cosine of the viewing angle is shown in Fig.~\ref{fig:wmu_ar_qs} for the AR dataset (top panel) and the QS dataset (bottom panel). There is, on average, a slightly larger line width at the limb than at disk center, for both datasets. The line broadening appears, on average, to increase by only a small amount towards the limb, of order 1-2 km/s. The increase is larger for the AR 12920 dataset.
The PDF of the line-core intensity (from the fit) as a function of viewing angle is shown in Fig.~\ref{fig:imu_ar_qs} for the two datasets. The intensity increases towards the limb for both datasets. For QS, the core intensity increases from an average of 0.3 DN/s at disk center to 2.7 DN/s at the limb. The corresponding numbers for the AR dataset is an increase from 0.48 DN/s at disk center to 3.7 DN/s. The AR dataset includes a larger variety of features and shows a larger spread in intensity.

We first discuss the center-to-limb behavior of the line broadening on the disk (i.e., not including the off-limb region). The changed viewing angle towards the limb could, in principle, cause two effects on the \oi\ line broadening. First, it could mean that the measurements closer to the limb are more sensitive to the motions perpendicular to the magnetic field. This would be the case if one assumes that, on average, the magnetic field is typically more vertically (rather than horizontally) oriented in the line formation region of \oi. This is likely a reasonable assumption for a plage or network region, but may not be a good assumption for other regions in the FOV, such as those regions just adjacent to network or plage where fibril or spicule-like features often appear more inclined, or for internetwork regions where the chromospheric field is poorly known and may include more horizontal fields. Secondly, the longer line-of-sight implies more superposition of different structures within each pixel. If these locations have different LOS velocities, this can broaden the line as well. If this were the case, one would also expect to see a significant increase in intensity, depending on the velocity gradients in the FOV and/or along the line-of-sight.

Disentangling these two effects is not straightforward. In addition, the likely different field orientations between network or plage on the one hand, and the other regions on the other hand, can easily render interpretation of a center to limb variation plot such as that in Fig.~\ref{fig:wmu_ar_qs} muddled. By mixing both types of regions into one plot, any center-to-limb variation of each sub-region may be hidden since the two different types of regions may have oppositely signed center-to-limb variations. Another key aspect that should be taken into account is that the line formation region of \oi\ is very large, covering a region from the low chromosphere all the way to the top of the chromospheric spicules. We have already discussed the presence of spicule-like features in the \oi\ maps of Figs.~\ref{fig:overview} and \ref{fig:overview_qs}. The likely field-aligned motions and the amplitude of, e.g., Alfv\'en-wave associated motions perpendicular to the field, are expected to vary very significantly between the dense lower chromosphere and the top of spicules, because of the very large density differences between those two regions. Assuming mass conservation for field-aligned flows, or constant energy flux density for Alfv\'en waves implies a strong increase of the flows from the low chromosphere to the top of spicules. In addition, there are most likely significant differences in intensity between the chromosphere and the top of spicules, since the intensity is proportional to the square of the electron density for this optically thin line \citep{Lin2015}. The relative contribution of these two components (chromosphere vs. spicules) within an IRIS resolution element is also expected to vary between disk center and the limb. This is because at disk center, spicules will more often originate and be suspended straight above the bright network or plage regions, whereas toward the limb their greater height will show them more spatially offset from the network or plage.

\begin{figure}[tph]
    \centering
    \includegraphics[width=0.45\textwidth]{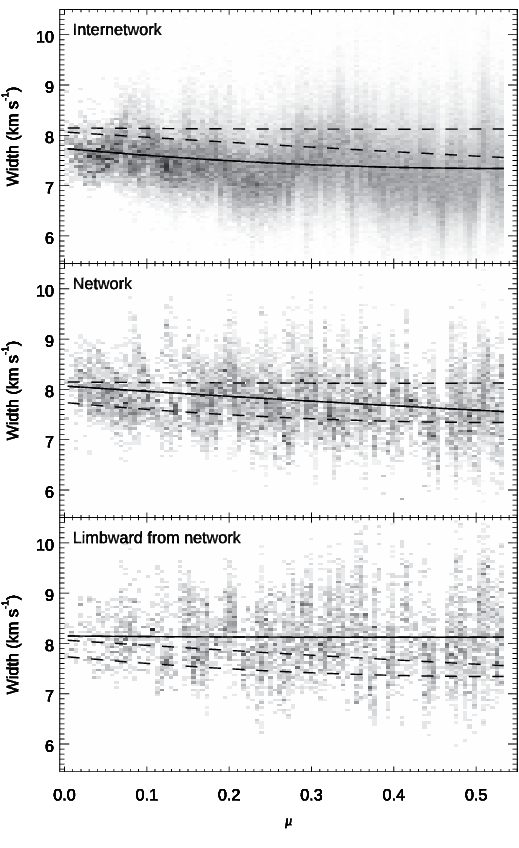}
    \caption{Probability density function (PDF) of the \oi\ line width as function of the cosine of the viewing angle ($\mu$) for quiet Sun internetwork areas (top panel), network areas (middle panel) and for areas 1\arcsec\ limbward of network areas (bottom panel). In all panels the least-squares quadratic fit is shown with a solid line and the fits for the two other area-types as dashed lines.}
    \label{fig:wmu_qs}
\end{figure}

In an attempt to disentangle these effects we have applied masks to isolate the chromospheric from the spicular contributions as much as possible. We perform this only for the quiet Sun dataset since the magnetic field in active regions is more complex. Fig.~\ref{fig:wmu_qs} shows the center-to-limb variation of quiet Sun network, regions adjacent to the network, and the internetwork. 
The masks are created by first making a least-squares cubic polynomial fit to the core intensity (from the fit) as a function of the cosine of the viewing angle ($\mu$). The network mask includes all pixels with a core intensity more than 22 DN above the center-to-limb fit. The "adjacent to the network" mask includes all pixels less than one arcsecond limb-ward from the network pixels, excluding network pixels. The internetwork mask includes all the other pixels.
The regions adjacent to the network are typically dominated by fibrils or spicules. We see that the regions adjacent to the network show an increased line broadening. Because of the morphology, we believe that this is caused by the larger motions in the low density environment of spicules. It is possible that this effect also contributes to the modest increase of the line width by 1-2 km~s$^{-1}$ increase towards the limb that is seen in Fig.~\ref{fig:wmu_ar_qs}. The fainter spicular signal (with its increased LOS motions) becomes more apparent towards the limb because the spatial offset from the brighter network is increased as the viewing angle changes. Another effect that may help explain the modest increase of line width toward the limb is the increased superposition along the line-of-sight of different structures with different LOS velocities. However, 
it is also possible that towards the limb the LOS is more perpendicular to the magnetic field direction, and that the modest increase of line width towards the limb is in part caused by stronger motions perpendicular to the magnetic field direction. This slight anisotropy (with respect to the magnetic field direction) of turbulent motions likely plays a significant role, as we illustrate in what follows below.

\begin{figure}[tph]
    \centering
    \includegraphics[width=0.45\textwidth]{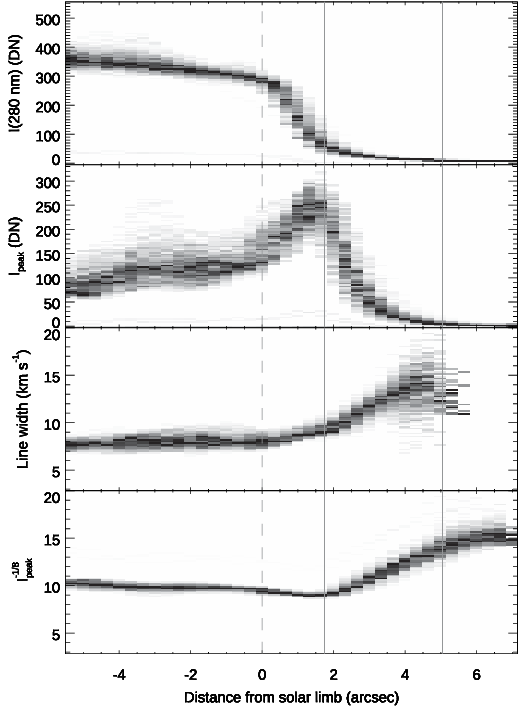}
    \caption{Probability density functions (PDF) for the intensity at 2800\AA\ (upper photosphere, top), \oi\ intensity $I_{peak}$ (second from top), \oi\ line width (third from top), and a quantity $A$ that is proportional to $I_{peak}^{-1 \over 8}$, all as a function of distance from solar limb, with positive values of the latter for off-limb positions. The vertical dashed line is the location of the photospheric limb, as determined from the solar radius and \iris\ headers. The solid vertical lines show the range of distances from the limb for which $I_{peak}$ decreases with height. The value of $A$ is scaled so that it is equal to the line broadening at the distance from the solar limb indicated by the leftmost solid vertical line. }
    \label{fig:mats4}
\end{figure}

We now turn our attention to the off-limb behavior of the \oi\ intensity and line broadening in a quiet Sun dataset as a function of distance above the limb (positive values). This is shown in Fig.~\ref{fig:mats4}.  The distance from the limb is based on a calculation of the solar radius for the date of the observation and the header information in the \iris\ data. The latter has an accuracy of order $\sim 0.6$\arcsec\ or better, now that cross-correlation of \iris\ FUV SJI images and AIA 1700\AA\ data is automatically applied in the \iris\ level 2 data pipeline. The top panel shows the sharp drop-off of the intensity in the photospheric wing of \mgiik\ at 2800\AA, which starts about 0.5\arcsec\ above the solar limb. This is a reasonable number since this photospheric wing emission is formed in the upper photosphere, confirming that the limb distance is accurately calculated.

The second panel from the top in Fig.~\ref{fig:mats4} shows that the \oi\ intensity increases from the photospheric limb outward until it peaks at about 1.5\arcsec from the limb. The intensity then rapidly drops off with increasing distance from the limb. It is very interesting to note that between the limb and the 1.5\arcsec\ distance where the \oi\ intensity peaks, the line width increases (for increasing distance from the limb) only very modestly to a value of about 9 km~s$^{-1}$. This changes drastically after the peak intensity is reached: the line width then increases rapidly to values of 15 km~s$^{-1}$ until the intensity drops (at distances of 5 \arcsec) to very low values where the Gaussian fits can no longer be performed reliably.

What causes this puzzling spatial offset of about 3\arcsec\ of the off-limb peak of the intensity and the line width? If one assumed that the off-limb properties of this optically thin line were caused by line-of-sight superposition of intensity features with a random distribution of LOS velocities, the peaks of the intensity and line width should be co-located.

Let us, instead, examine a different scenario that is inspired by our findings in \S~\ref{stats}. We assume that there are two major contributors to the \oi\ signals off-limb: a contribution from the chromosphere proper, and one from spicules. Let us further assume that, at the limb, the line width is mostly determined by motions perpendicular to the magnetic field, as expected from Alfv\'en wave associated motions, and that the field is mostly vertical. As we have seen in \S~\ref{stats}, there are indications that the line width increases along the spicule-like structures emanating from the network. It is natural to assume that beyond a certain height, these spicular features dominate the signal. 

Let us now assume that the energy flux density $F$ of Alfv\'en waves propagating from the lower chromosphere to the top of the spicules is roughly constant:
\begin{equation}
F = \rho \delta v^2 v_A \sim \sqrt{\rho} \delta v^2 B = k
\label{eq1}
    \end{equation}

in which $\rho$ is the mass density of the plasma, $\delta v$ is the Alfv\'en wave amplitude,  $v_A$ is the Alfv\'en speed, with $v_A \sim B / \sqrt{\rho}$, and $B$ the magnetic field, and $k$ a constant. This would then imply:
\begin{equation}
\delta v \sim \rho^{- {1 \over 4}} {B}^{- {1 \over 2}} \sim n_e^{-{1 \over 4}} {B}^{- {1 \over 2}}  \sim I_{peak}^{-{1 \over 8}} {B}^{- {1 \over 2}} 
\end{equation}
in which we assume that $\rho \sim n_e$ (the electron density), and $I_{peak} \sim n_e^2$ \citep[based on][]{Lin2015}.

In the bottom panel of Fig.~\ref{fig:mats4}, we show the variation of the component $I^{-{1 \over 8}}$, normalized to the same line width values at the distance where the \oi\ line width peaks (about 1.5\arcsec). We find that the increase of line width from 1.5\arcsec\ to 5\arcsec is well reproduced qualitatively. Note that we expect that the average magnetic field decreases with increasing distance from the limb. Such a decrease in magnetic field $B$ would lead to an even larger increase of the line width. It seems that our relatively simplistic model thus reproduces the observed behavior very well. Note also that our approach appears to also be compatible with the observed slight increase of line width between the solar limb and the distance of peak \oi\ intensity. While the $I_{peak}^{-1 \over 8}$ component predicts a slight decrease in line width in this region, we note that it is precisely here (at heights between the photosphere and low chromosphere) that we expect the largest drop in average magnetic field strength as flux tubes expand with height from the photosphere into the low chromosphere. If the field were to drop by a factor of 3, this could reverse the predicted decrease of line width (from the intensity alone) to the modest increase we actually see for increasing distance from the solar limb.

Our results thus support a scenario in which the spatial variation of the \oi\ linewidth is related to the superposition of Alfv\'enic wave motions or turbulence along  structures that show a density that significantly decreases with height. Naturally reality is most likely more complex. However, the fact that this scenario can easily explain the spatial offset between the peaks of \oi\ intensity and line width, and shows an increase of line width that is self-consistent with the observed intensities and expected magnetic field variations, both suggest that this scenario plays a role in the observed behavior. 

\begin{figure*}
    \centering
    \includegraphics[width=0.99\textwidth]{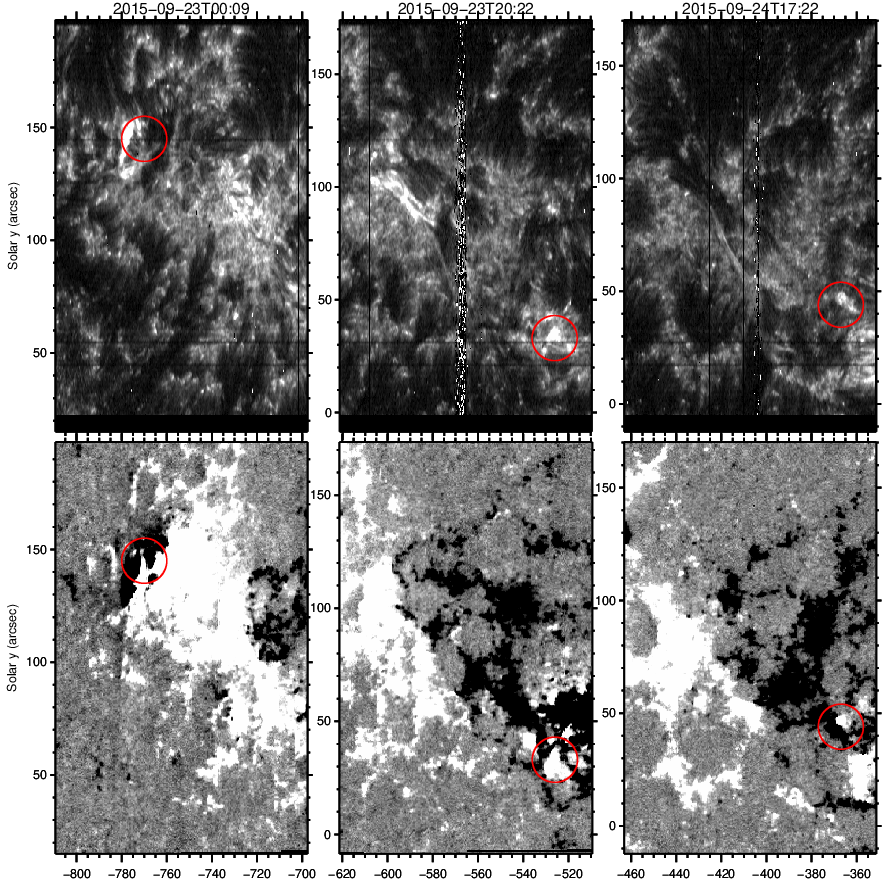}
    \caption{\iris\ spectroheliograms of NOAA AR 12920 as it traverses the disk between 2015-09-22 and 2015-09-24. Top row shows the  \ion{O}{1} 1355\AA\ intensity, while the bottom row shows the line-of-sight magnetic field as deduced from \hmi\ magnetograms stitched together to replicate the \iris\ raster timings and locations. The red circles indicate locations with neutral lines between opposite polarities and increased \oi\ intensities, as described in the text.}
    \label{fig:hmi1}
\end{figure*}

\begin{figure*}
    \centering
    \includegraphics[width=0.99\textwidth]{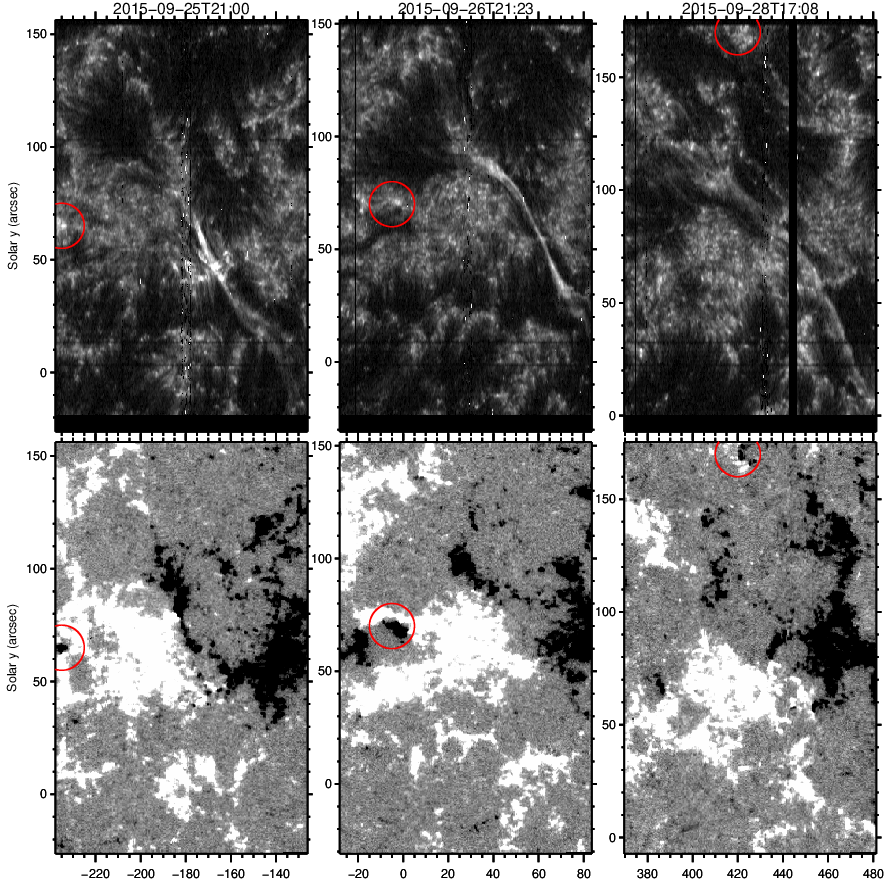}
    \caption{\iris\ spectroheliograms of NOAA AR 12920 as it traverses the disk between 2015-09-25 and 2015-09-28. Top row shows the  \ion{O}{1} 1355\AA\ intensity, while the bottom row shows the line-of-sight magnetic field as deduced from \hmi\ magnetograms stitched together to replicate the \iris\ raster timings and locations. The red circles indicate locations with neutral lines between opposite polarities and increased \oi\ intensities, as described in the text.}
    \label{fig:hmi2}
\end{figure*}

\section{Relationship to magnetic flux emergence and cancellation}\label{cancel}

The very bright features in \oi\ in the active region (Fig.~\ref{fig:c2l_ar}) appear to be related to the effects of flux emergence and/or cancellation. A detailed comparison of the locations in \oi\ and magnetograms from \hmi\ (Fig.~\ref{fig:hmi1}, \ref{fig:hmi2}) shows that the \oi\ intensity is often enhanced around the neutral line between opposite polarities. To allow a proper comparison between the \iris\ rasters and \hmi\ (or \sst) magnetograms, we create, from the \hmi\ (or \sst) data, synthetic rasters with identical field-of-view as the \iris\ raster, in which each raster position includes a vertical strip of \hmi\ (or \sst) data at the same time as the time of each \iris\ raster position.

On the largest scales ($\sim 20$\arcsec), we see that filaments, which most often occur above the neutral line between opposite polarities, are often bright in \oi\, e.g., at (-166\arcsec, 42\arcsec) (left), and at (53\arcsec, 70\arcsec) (middle) in Fig.~\ref{fig:hmi2}. However, the correlation between \oi\ intensity and neutral lines is also strong on smaller scales ($\sim 5$\arcsec), at locations where flux concentrations of opposite polarities are in very close proximity, i.e., touching or almost touching. This is clearly illustrated, in Fig.~\ref{fig:hmi1} (red circles), by the bright region at (-770\arcsec, 145\arcsec) in the left column, at (-526\arcsec, 33\arcsec) in the middle column, and (-367\arcsec, 44\arcsec) in the right column. Similarly, Fig.~\ref{fig:hmi2} (red circles) shows examples at (-235\arcsec, 65\arcsec) (left), (-5\arcsec, 70\arcsec) (middle), and (420\arcsec, 170\arcsec) (right). 

\begin{figure*}
    \centering
    \includegraphics[width=0.99\textwidth]{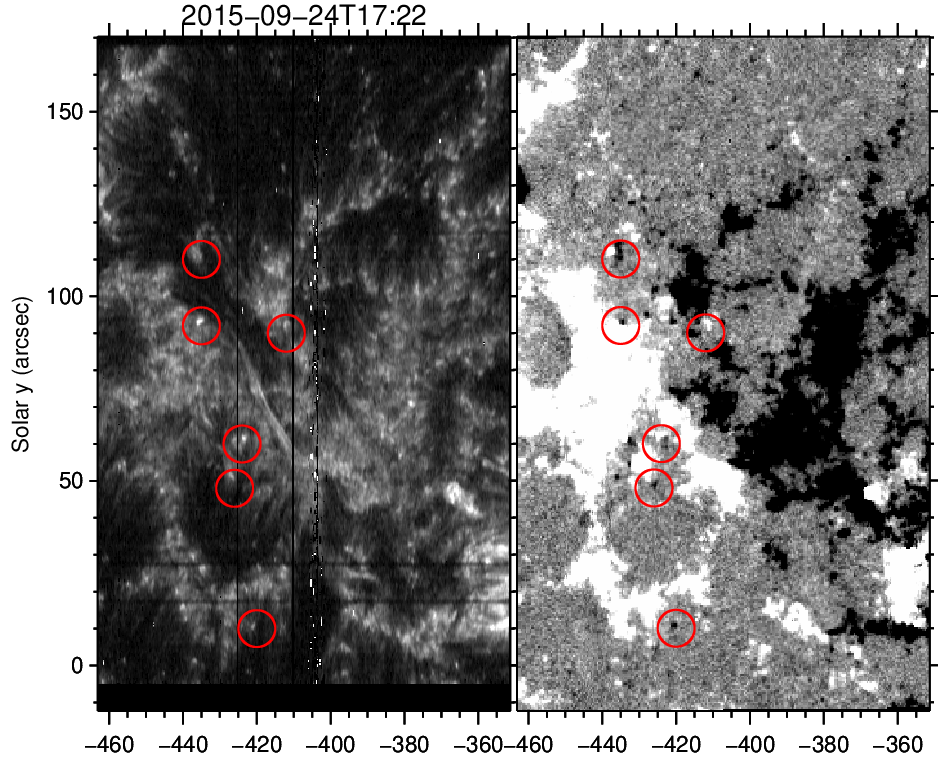}
    \caption{\iris\ spectroheliograms of NOAA AR 12920 on 2015-09-24. Left column shows the  \ion{O}{1} 1355\AA\ intensity, while the right column row shows the line-of-sight magnetic field as deduced from \hmi\ magnetograms stitched together to replicate the \iris\ raster timings and locations. The red circles indicate locations with neutral lines between opposite polarities and increased \oi\ intensities, as described in the text.}
    \label{fig:hmi3}
\end{figure*}

When we saturate \hmi\ magnetograms to enhance the visibility of weak flux of opposite polarity, we find that on arsecond scales very weak flux immediately adjacent to the dominant polarity flux similarly often leads to enhanced brightenings in \oi. Examples of that can be found in Fig.~\ref{fig:hmi3} (red circles), at (-435\arcsec, 110\arcsec), (-435\arcsec, 92\arcsec), (-412\arcsec, 90\arcsec), (-424\arcsec, 60\arcsec), (-426\arcsec, 48\arcsec), (-420\arcsec, 10\arcsec). There are, however, also some locations in which opposite polarity flux appears in close proximity but the \oi\ intensity does not appear to be significantly enhanced (e.g., the weak concentrations at (-370\arcsec, 55\arcsec)).

\begin{figure*}
    \centering
    \includegraphics[width=0.99\textwidth]{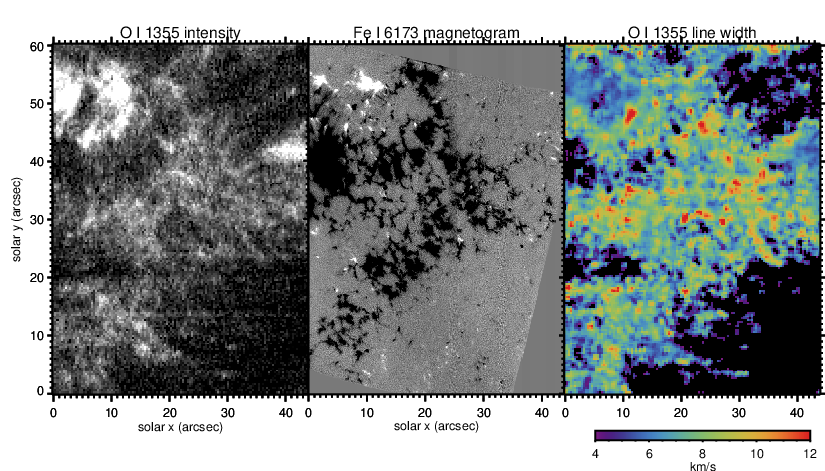}
    \caption{Comparison between SST Fe I 6173\AA\ magnetogram raster (middle) and \iris\ \ion{O}{1} intensity (left) and line width (right), taken on 2015-09-09. The magnetogram synthetic raster has been constructed from a magnetogram timeseries by matching with \iris\ raster timings and locations.
    The \oi\ intensity is scaled between 2 and 30 DN. Regions with bad fits and/or low peak counts (below 5 DN) are masked out in the \oi\ line width map.
    This figure is accompanied by an animation with the same layout as the figure, except it shows in the middle panel a time sequence of magnetograms. Short vertical bars at the top and bottom of each panel in the animation indicate, for each time step, the location of the \iris\ raster step (left and right panels) at the time of the magnetogram shown in the middle panel. The time in the animation is expressed in seconds after 9-Sep-2015 09:00:00 UTC. }
    \label{fig:sst2}
\end{figure*}

To investigate this correlation on sub-arcsecond scales, we also analyze a time series of very high-resolution magnetograms obtained at the SST (Fig.~\ref{fig:sst2}). These magnetograms have much higher spatial resolution ($\sim 0.1\arcsec$) and greater magnetic sensitivity. They similarly reveal locations of enhanced \oi\ brightenings where flux of the minority polarity is in the proximity of stronger dominant polarity concentrations, e.g., at (4\arcsec, 17\arcsec). However, there are also locations with very weak minority flux where this correlation is not as clear (e.g., at (12\arcsec, 35\arcsec)) or not apparent (e.g., at (18\arcsec, 56 \arcsec)). Such locations of reduced or no correlation appear to be more common in the SST dataset than in the HMI datasets. This perhaps suggests that a minimum flux size is required for significant \oi\ emission, or that the \oi\ intensity increase is perhaps shorter lived for very weak minority flux concentrations, and missed by the slow cadence of the \iris\ rasters. The latter is driven by the required deep exposures to detect the faint \oi\ line.

The animation associated with Fig.~\ref{fig:sst2} reveals that, in addition to the correlation between neutral lines and \oi\ intensity, which could be associated with flux cancellation, there is another process that is clearly associated with increased \oi\ intensity. Flux emergence is present in the region around $x=5-15\arcsec$, $y=45-55\arcsec$, and $x=38-43\arcsec$, $y=40-45\arcsec$. These are the two regions that show the brightest \oi\ emission. This can also be seen in Fig.~\ref{fig:ar_blink}, with the brightest emission around (-465\arcsec, -210\arcsec) occurring at a location where flux has emerged into a pre-existing field configuration. The region of enhanced intensities in \oi\ (also seen in \ion{Si}{4} and \ion{Cl}{1}) appears to outline a dome-like structure that separates the two flux systems, possibly a quasi-separatrix layer (QSL).

The above findings appear to be compatible with a scenario in which significant ionization and heating of the plasma, in response to reconnection or currents associated with the interaction between field concentrations of opposite polarity, leads to enhanced electron densities, which, based on theoretical work, the \oi\ intensity is proportional to \citep{Lin2015}.  Our observational results indicate that such heating could occur in association with flux cancellation or flux emergence. Out of all \iris\ observables the \oi\ intensity appears to be the most sensitive to the effects of cancellation and emergence, possibly because it is formed at low enough heights that it is sensitive to heating, even from small-scale flux concentrations whose fields do not reach into the upper chromosphere. Another key aspect is that \oi\ is optically thin and uniquely sensitive to the electron density, thereby picking up any enhancement caused by heating or ionization. It appears that \oi\ may be acting as a canary in the coal mine for solar atmospheric effects of cancellation or emergence.

One complication of our analysis is that the \oi\ line is faint and requires deep exposures, leading to low cadence observations. Given the dynamic and ephemeral nature of heating associated with flux emergence it is thus quite possible that some heating signals are simply missed in the \oi\ rasters, when it occurs either before or after the \iris\ slit has passed the site of cancellation or emergence. Alternatively, it may be that not all cancellations or emergence lead to significant heating at chromospheric heights. Detailed comparisons with numerical simulations are required to address this.

One final observational finding is that the locations of enhanced line broadening do not appear to show a significant correlation with flux cancellation or emergence, as illustrated by 
Fig.~\ref{fig:sst2}. In fact, regions of active emergence often appear to show significantly reduced broadening (e.g., at (5\arcsec, 52\arcsec) in Fig.~\ref{fig:sst2}).

\section{Flux concentrations} \label{tube}

\begin{figure*}
    \centering
    \includegraphics[width=0.9\textwidth]{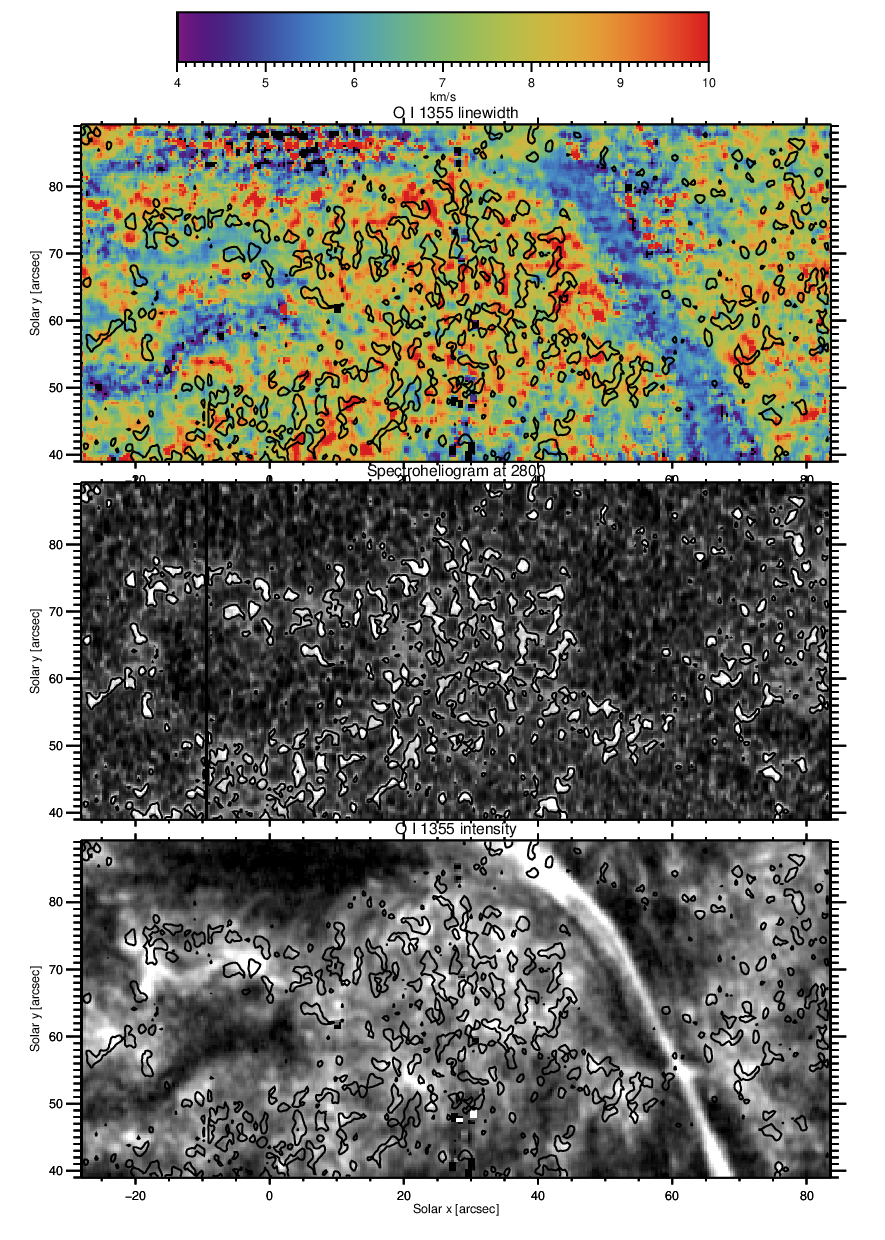}
    \caption{For NOAA AR 12902 on 2015-09-26, the \oi\ line width in km/s, the intensity at 2800\AA\ (formed in the photosphere), and the \ion{O}{1} 1355\AA\ intensity. Black contours are based on intensity thresholding of photospheric bright points in the 2800\AA\ spectoheliogram. The \oi\ intensity is scaled from 3 to 40 DN.}
    \label{fig:fluxtubes}
\end{figure*}

Our analysis of the \oi\ measurements and the underlying photosphere in plage regions has revealed another intriguing finding. In particular, on sub-arcsecond spatial scales, there is an anti-correlation between locations of enhanced line broadening and the locations of strong flux concentrations in the photosphere. This is illustrated in Fig.~\ref{fig:fluxtubes}, which shows the \oi\ line broadening (top), a spectroheliogram at 2800\AA\ (middle), and the \oi\ intensity (bottom). 

While the range of values for line broadening is not enormous, there is a clear difference between non-plage regions and plage regions, with the former (e.g., $x=-10$ to 15\arcsec, $y=80$ to 90\arcsec), showing significantly reduced line broadening (less than 5 km/s). The plage region itself shows a relatively narrow range of values for the line broadening, between 7 and 11 km/s, as already remarked upon by \citet{Carlsson:2015fk}. The top panel of Fig.~\ref{fig:fluxtubes} nevertheless reveals a spatial pattern of enhanced values for the line broadening (3 to 4 km/s higher than the rest of the plage), occurring in contiguous regions showing coherence on 0.5 to 2\arcsec spatial scales. These regions of enhanced line width do not seem to correlate with locations of enhanced \oi\ brightness. Similar, they do not occur where the photospheric wing emission at 2800\AA\ occurs. Instead, they preferentially avoid the locations of bright points in the photosphere. The latter are indicated with black contours, which are determined by setting a threshold brightness in the 2800\AA\ spectroheliogram. As can be seen in the top panel, most often the regions of enhanced line broadening in \oi\ occur in between the black contours. 

\begin{figure}
    \centering
    \includegraphics[width=0.48\textwidth]{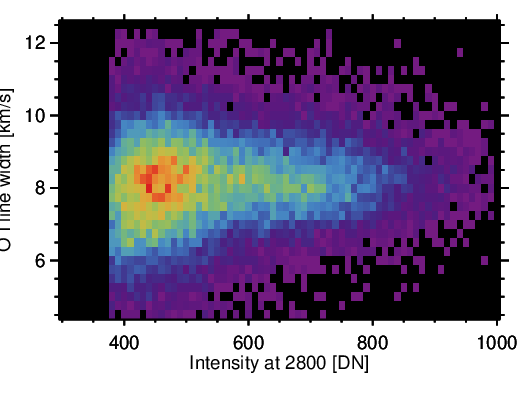}
    \includegraphics[width=0.48\textwidth]{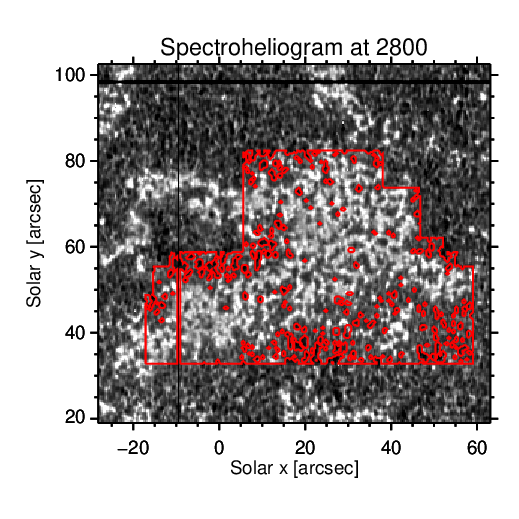}
    \caption{For NOAA AR 12902 on 2015-09-26, the JPDF (top panel) between the \oi\ line width in km/s, and the intensity at 2800\AA\ (formed in the photosphere) for a subset of the plage region studied in Fig.~\ref{fig:fluxtubes}. The JPDF only includes locations for which the \oi\ intensity is larger than 10 DN (to avoid noisy profiles with uncertain or inaccurate line width) and for locations that are within the red contours shown in the spectroheliogram at 2800 shown in the bottom panel.}
    \label{fig:fluxtubes2}
\end{figure}

This anti-correlation is further illustrated by Fig.~\ref{fig:fluxtubes2}, which shows the joint probability density function (JPDF) of the \oi\ line broadening and the intensity at 2800\AA. Only the locations where the \oi\ intensity is larger than 10 DN are included, within the plage region that is within the red contours in the bottom panel of Fig.~\ref{fig:fluxtubes2}. By introducing the threshold for \oi\ intensity we ensure that the resulting line widths included in the JPDF are not significantly affected by noise. 

As we can see, the locations of highest broadening occur at lower values in photospheric intensity. The locations with highest photospheric intensity (bright points) show moderate levels of broadening. The lowest broadening occurs at the edge of the plage regions, as seen in the top panel of Fig.~\ref{fig:fluxtubes}. The scatter plot thus supports our finding that the locations of highest line broadening occur in locations with the lowest values of 2800\AA\ intensity within plage regions, i.e., those in between photospheric bright points. We note that the broader range of values for the \oi\ broadening at lower 2800\AA\ intensities is not a signature of noise, as the noise in determining the line width depends on the \oi\ intensity (not the photospheric intensity). If noise were behind this issue, then we would expect a correlation between decreased \oi\ intensity and decreased photospheric intensity. That is not the case, as can be seen in the bottom two panels of Fig.~\ref{fig:fluxtubes}.

It is also noteworthy that the \oi\ line broadening map does not show a good correlation or anti-correlation with any other width or intensity measure of other chromospheric or TR lines. There are perhaps a few locations where there is some correspondence with \mgiik\ line broadening, but this is not the general trend (see Figs.~\ref{fig:overview} and ~\ref{fig:overview_qs}). 

This begs the question what causes this anti-correlation. Several scenarios come to mind. First, the area between photospheric bright points in plage is the region where the magnetic field canopy is expected to occur, as seen in both observations \citep[e.g.,][]{de-la-Cruz-Rodriguez2019} and numerical simulations \citep[e.g.,][]{Hansteen+DePontieu2006}. While the magnetic field is expected to be mostly vertical directly above photospheric flux concentrations, it is expected to be significantly more inclined with respect to the vertical in the regions where the canopy forms, i.e., the locations where we find the highest \oi\ line broadening. This is particularly the case at low chromospheric heights where the transition occurs from high to low plasma $\beta$. The \oi\ line is thought to be sensitive to the electron density, with the formation height covering the low chromospheric regions in particular. 

But why would a heavily inclined magnetic field lead to enhanced \oi\ line broadening? The line width is determined by three contributors. The instrumental broadening, which does not vary significantly across the FOV; the thermal broadening, which depends on the local temperature, and the unresolved motions along the line-of-sight ("non-thermal" broadening). If increased heating (i.e., higher temperatures) were to occur in the chromospheric canopy region between photospheric bright points, it would be expected to lead to enhanced electron densities. This in turn would cause enhanced \oi\ intensities, since this optically thin line scales with the square of the electron density. However, such a correlation is not observed in our data (top and bottom panels of Fig.~\ref{fig:fluxtubes}). 

This leaves the possibility that enhanced non-thermal motions, either in the form of unresolved macroscopic motions along the line-of-sight, or in the form of microscopic turbulence, are the cause of these enhancements in line broadening. It is tempting to speculate that this may be caused by one (or both) of two scenarios. The first is one in which strong Alfv\'enic wave and/or vortical motions (i.e., perpendicular to the magnetic field direction) are ubiquitous in plage and register in the \oi\ line broadening only when the magnetic field direction is more inclined from the line-of-sight vector. In a disk center observation like the one considered here, such enhanced vortical motions or Alfv\'enic wave power would thus be most visible in between photospheric bright points. This is an intriguing scenario given the evidence from the center-to-limb variation of line broadening that Alfv\'enic waves play a key role in explaining the strong increase of line broadening off-limb. These observational findings may also be compatible with recent suggestions from numerical simulations that vortical motions in the photosphere are ubiquitous and often propagate into the low atmosphere \citep[e.g.,][]{Moll2011, Yadav2021, Breu2023}. However, it is not fully clear whether these recent modeling results are fully compatible with our observations. In particular, the lack of correlation between locations of enhanced line width and \oi\ intensity in observations does, at first blush, not seem to be fully compatible with a scenario in which the vortices lead to increased heating in the plage chromosphere \citep{Yadav2021}. This is because one would expect that such heating leads to enhanced \oi\ densities and thus intensities. Detailed studies of synthetic \oi\ profiles are required to further investigate this. 

An alternative possibility is that there are strong velocity gradients at the interface of flux concentrations. These could, for example, occur because of the LOS overlap between upward propagating shocks on neighboring flux concentrations. Such shocks are ubiquitous in plage and drive strong flows along dynamic fibrils \citep{Hansteen+DePontieu2006}. Perhaps a combination of both effects plays a role. Detailed comparisons with advanced numerical simulations of the chromosphere are required to further investigate these scenarios.

\section{Conclusions}\label{dis}

We have analyzed the properties of the \oi\ spectral line, which is unique in that it is regularly observed with \iris, optically thin, and formed in the chromosphere. We find that the line shows properties that are different from other spectral lines formed in the chromosphere. We find that intensities are strongest in plage and network regions and in the proximity of neutral lines where magnetic fields of opposite polarity are in close contact. Our data suggests that the \oi\ intensity is often increased and appears to be very sensitive to the effects of cancellation and emergence of magnetic flux. Because of the optically thin nature of the \oi\ line, this indicates that electron densities are locally enhanced, a signature of heating in the chromosphere. We also see a significant increase of \oi\ intensities at the solar limb, as expected for an optically thin spectral line from the increased line-of-sight superposition of different structures. We find evidence for \oi\ intensity structures being associated with shocks in plage and network, as well as spicules, with some similarities to counterparts at low TR temperatures that are visible in \ion{Si}{4}lines.

The \oi\ line width offers a unique view on the unresolved or non-thermal motions in the chromosphere, a quantity that is otherwise difficult to directly determine, requiring inversions of optically thick spectral lines whose formation is subject to non-LTE radiative transfer effects. We find, for both active regions and quiet Sun, that the line width modestly increases towards the limb, and along spicule-like structures that protrude away from plage and network flux concentrations. The modest center-limb-variation suggests that there are unresolved motions both along the magnetic field and across the magnetic field, with the latter being stronger.

Off the limb the line broadening rapidly increases, compatible with a scenario in which turbulent, Alfv\'en wave or vortical motions perpendicular to the magnetic field dominate the line width, as they propagate upward with a roughly constant flux along spicular structures in which the density decreases with height. 

The presence of strong vortical or Alfv\'enic motions in plage are further supported by a curious but significant enhancement of line width in between photospheric flux concentrations, when viewed close to disk center. These results are compatible with the combined impact of inclined canopy fields and enhanced motions perpendicular to the magnetic field on the \oi\ linewidth. Such inclined fields are expected to occur in between photospheric flux concentrations, while enhanced motions perpendicular to the field are predicted by various numerical simulations of vortices and Alfv\'en waves in plage. However, the lack of obvious correlation between enhanced line width and \oi\ intensity is not immediately compatible with the enhanced heating expected from vortices.

Our observations provide strict constraints on models of the chromosphere.

\acknowledgements{\longacknowledgment} 

\bibliographystyle{aasjournal}
\bibliography{senior22,collectionbib,Bibliography}

\end{document}